\documentclass[a4paper,12pt]{article}
\usepackage[pctex32]{graphics}

\begin{document}

\topmargin 0pt \oddsidemargin 0mm
\newcommand{\be}{\begin{equation}}
\newcommand{\ee}{\end{equation}}
\newcommand{\ba}{\begin{eqnarray}}
\newcommand{\ea}{\end{eqnarray}}
\newcommand{\fr}{\frac}

\begin{titlepage}
\begin{flushright}
{\tt gr-qc/0611130}
\end{flushright}

\vspace{5mm}
\begin{center}
{\Large \bf Thermodynamics and evaporation of the noncommutative
black hole } \vspace{12mm}

{\large   Yun Soo Myung$^{\rm a,}$\footnote{e-mail
 address: ysmyung@inje.ac.kr},
 Yong-Wan Kim $^{\rm b,}$\footnote{e-mail
 address: ywkim@pcu.ac.kr},
and Young-Jai Park$^{\rm c,}$\footnote{e-mail
 address: yjpark@sogang.ac.kr}}
 \\
\vspace{10mm} {\em $^{\rm a}$Institute of Mathematical Science and
School of Computer Aided Science, \\Inje University, Gimhae 621-749,
Korea \\} {\em $^{\rm b}$ National Creative Research Initiative
Center for Controlling Optical Chaos, \\Pai-Chai University, Daejeon
302-735, Korea \\} {\em $^{\rm c}$Department of Physics  and Center
for Quantum Spacetime,\\ Sogang University, Seoul 121-742, Korea}
\end{center}

\vspace{5mm}

\centerline{{\bf{Abstract}}}

\vspace{5mm} We investigate the thermodynamics of the
noncommutative black hole whose static picture is similar to that
of the nonsingular  black hole known as the de
Sitter-Schwarzschild black hole. It turns out that the final
remnant of extremal black hole is a thermodynamically stable
object. We describe the evaporation process of this black hole by
using the noncommutativity-corrected Vaidya metric. It is found
that there exists a close relationship between thermodynamic
approach and evaporation process.

\vspace{3mm}

\end{titlepage}

\newpage

\renewcommand{\thefootnote}{\arabic{footnote}}
\setcounter{footnote}{0} \setcounter{page}{2}

\section{Introduction}
Hawking's semiclassical analysis of  the black hole  radiation
suggests that most  information about initial states is shielded
behind the event horizon and will not back to the asymptotic region
far from the evaporating black hole~\cite{HAW1}. This means that the
unitarity is violated by an evaporating black hole. However, this
conclusion has been debated  by many authors for three
decades~\cite{THOO,SUS,PAG}. It is closely related  to the
information loss paradox, which states the question of whether the
formation and subsequent evaporation
 of a black hole is unitary. One of the most urgent problems in the black
hole physics is  the lack of resolution of  the unitarity issue.
Moreover, a complete description of the final stage of the black
hole evaporation is important but is still quite unknown. In order
to reach the solution to these problems, we have to use quantum
gravity. Although two leading candidates for quantum gravity are
the string theory and the loop quantum gravity, we need to
introduce another approach that provides a manageable form of the
quantum gravity effect. The holographic principle could serve such
a purpose because it includes the effect of the quantum mechanics
and gravity~\cite{Ho,Su}.

Also it is interesting to consider  the generalized uncertainty
principle (GUP)  since the Heisenberg uncertainty principle may
not be satisfied when quantum gravitational effects become
important~\cite{nc,GUP,ACS,mo}. We note that  the GUP
provides the minimal length scale and thus modifies the
thermodynamics of a singular black hole at the Planck scale only.

On the other hand, even though the noncommutativity  also provides
the minimal length scale~\cite{nc},  this provides a totally
different black hole: noncommutative black hole (NBH)~\cite{SS}.
This is similar to  the nonsingular black hole with two
horizons~\cite{Dymn}. We think that  it is very important to study
the effects of noncommutativity on the terminal phase of black
hole evaporation. In case of the Schwarzschild black hole, the
temperature diverges and a large curvature state is reached.
However, it was shown that the noncommutativity can cure this
pathological short distance behavior~\cite{nicol}.

In this work, we first study thermodynamic properties of the NBH
thoroughly and then investigate its evaporation process.
Especially, we wish to point out the connection between
thermodynamic approach and evaporation process.

 The organization of this work is as follows.
We  study the  thermodynamics of the NBH in section II. Section
III is devoted to investigation of the evaporation process of the
NBH by introducing the noncommutativity-corrected Vaidya metric.
Finally, we discuss and summarize our results in section IV.

\section{Thermodynamics of noncommutative black hole}

It was shown that noncommutativity eliminates point-like
structures in favor of smeared objects in flat
spacetime~\cite{SS}. The effect of smearing is implemented by
substituting  the  Dirac-delta function  with Gaussian
distribution of the width $\sqrt{\theta}$. This is a coordinate
coherent approach to the noncommutativity  with planck units of
$c=\hbar=G=1$.   For this purpose, the mass density is chosen as

\begin{equation} \label{2eq1}
\rho_\theta(r)=\frac{M}{(4\pi\theta)^{\frac{3}{2}}}\exp(-\frac{r^2}{4\theta}),
\end{equation}
which plays the role of a matter source, and the total mass $M$ of
the source is diffused throughout a region of linear size
$\sqrt{\theta}$. $\theta$ comes from  the noncommutator of $[
x^\mu,x^\nu]=i\theta^{\mu\nu}$ with $\theta^{\mu\nu}=\theta~ {\rm
diag}[\epsilon_{1},\cdots,\epsilon_{D/2}]$ \cite{nic,bar,gem,FH}.
We note that the constancy of $\theta$ leads to a consistent
treatment of Lorentz invariance and unitarity~\cite{ss2}.
Eq.(\ref{2eq1}) provides a self-gravitating droplet of anisotropic
fluid whose energy-momentum tensor is given by $T^\mu_{~~\nu}={\rm
diag}[-\rho_\theta,p_r,p_\bot,p_\bot]$ with the radial pressure
$p_r=-\rho_\theta$ and tangential pressure $p_\bot=-\rho_\theta -
{\frac{1}{2}r}\partial_{r}\rho_\theta$. Solving the Einstein
equation $G_{\mu\nu}=8 \pi  T_{\mu\nu}$ leads to the solution

\begin{equation} \label{2eq2}
ds^2 \equiv g_{\mu\nu} dx^\mu dx^\nu= -F(r)dt^2+
F(r)^{-1}dr^2+r^2d\Omega^2_2.
\end{equation}
Here the metric function $F(r)$ is given by

\begin{equation} \label{2eq3}
F(r)=\Bigg[1-\frac{4M}{r\sqrt{\pi}}\gamma\Big(\frac{3}{2},\frac{r^2}{4\theta}\Big)\Bigg],
\end{equation}
\begin{figure}[t!]
   \centering
   \includegraphics{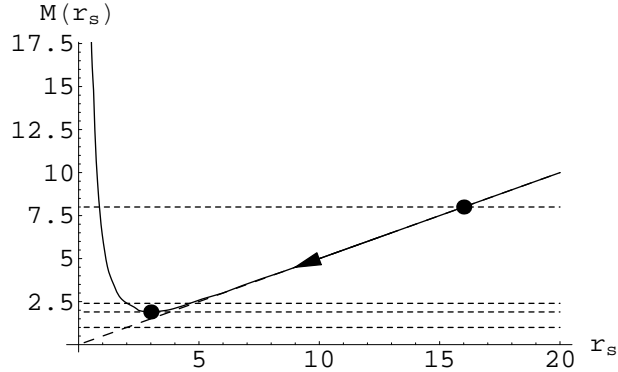}
\caption{The solid line: mass $M(r_s)$ as a function of the black
hole radius $r_s$. The dashed line is the Schwarzschild case. Four
horizontal lines are $M=8.0(=M_i), 2.4(=M_m), 1.9(=M_0),$ and
$1.0(<M_0)$ from top to bottom. The mass $M_i$ is introduced to be
the initial mass ($\bullet$) and $M_0(\bullet)$ is  the end point
for an  evaporation process ($\to$). For $M \ge M_0$, $r_C (\le
r_0)$ describes the inner horizon, while $r_s (\ge r_0)$
represents the outer horizon. de Sitter space appears for $0 \le
r_s<r_C$.} \label{fig1}
\end{figure}
\noindent where the lower incomplete gamma function is defined  by

\begin{equation} \label{2eq4}
\gamma\Big(\frac{3}{2},\frac{r^2}{4\theta}\Big)\equiv\int_0^{\frac{r^2}{4\theta}}t^{\frac{1}{2}}e^{-t}dt.
\end{equation}

\noindent Note that when $r$ goes to infinity, $\gamma$ approaches
to $\sqrt{\pi}/2$. From the condition of $g_{00} ( r_s ) = 0$, the
event horizon can be found as

\begin{equation} \label{2eq5}
r_s=\frac{4M}{\sqrt{\pi}}\gamma\Big(\frac{3}{2},\frac{r_s^2}{4\theta}\Big)\equiv \frac{4M}{\sqrt{\pi}}\gamma_s
\end{equation}
which provides the mass $M(r_s)$ as a function of the horizon
radius $r_s$
\begin{equation} \label{2eq6}
M(r_s)=\frac{\sqrt{\pi}r_s}{4\gamma_s}.
\end{equation}

\noindent In the large radius of $r_s^2/4\theta \gg1$(Hawking
regime: $r_{s}\simeq 2M$), the effect of noncommutativity can be
neglected. On the other hand, at the short distance of
$r_s^2/4\theta \simeq {\cal O}(1)$ (critical regime), one expects
to find significant changes due to the spacetime noncommutativity.
We note that although the  metric in Eq. (\ref{2eq2}) gives rise
to asymptotically Schwarzschild spacetime, it is basically
different from the Schwarzschild solution. This solution has two
parameters $M$ and $\theta$, in compared with one parameter $M$
for the Schwarzschild case.   Hereafter we choose $\theta=1$ for
numerical computations without any loss of generality. As is shown
Fig. 1, two masses of $M(r_s)$ and $M_{Sch}=r_s/2$ are  different
at the critical regime but the two are the same in the Hawking
regime. A minimum mass of $M(r_0)=M_0=1.9\sqrt{\theta}$ is
determined from the condition of $dM/dr_s=0$.

For definiteness,  we consider three different types: 1) For
$M>M_0(M=M_i=8.0\sqrt{\theta})$, two distinct horizons appear with
the inner (Cauchy) horizon $r_C$ and the outer (event) horizon
$r_s (r_C \le r_0 \le r_s)$. 2) In case of $M=M_0$, one has the
degenerate horizon at $r_0=3.0 \sqrt{\theta}$, which corresponds
to the extremal black hole. 3) For $M<M_0(M=1.0\sqrt{\theta})$,
there is no horizon. In case of $M \gg M_0$, the inner horizon
shrinks to zero, while the outer horizon approaches the
Schwarzschild radius $r_s=2M$. Hence the noncommutative black hole
solution looks like the nonsingular solution known as the de
Sitter-Schwarzschild black hole\footnote{At this point, one does
not confuse this noncommutative black hole with the
Schwarzschild-de Sitter black hole with $r_s \le r_C$, which
corresponds to a singular black hole inside the cosmological
horizon~\cite{GH}.}~\cite{Dymn}. Here $\rho_\theta$ connects the
de Sitter vacuum in the origin with the Minkowski vacuum at
infinity.

The black hole temperature in the noncommutative geometry can be
calculated
 to be \begin{equation} \label{2eq7}
T_{NBH}(r_s)\equiv
-\frac{1}{4\pi}\Bigg[\frac{dg_{00}}{dr}\Bigg]_{r=r_s}=\frac{1}{4\pi
r_s}\Bigg[1-\frac{r_s^3}{4\theta^{\frac{3}{2}}}\frac{e^{-\frac{r_s^2}{4\theta}}}
{\gamma_s}\Bigg].
\end{equation}

\noindent For $r_s^2/4\theta
 \gg 1$, one recovers the  Hawking temperature of the Schwarzschild
 black hole
\begin{equation}
T_H=\frac{1}{4\pi r_s}.
\end{equation}
Therefore, at the initial stage of Hawking radiation, the black
hole temperature increases as  the horizon radius decreases. It is
important to investigate what happens as $r_s
\rightarrow\sqrt{\theta}$. In the commutative case, $T_H$ diverges
and this puts limit on the validity of the conventional
description of Hawking radiation. Against this scenario, the
temperature $T_{NBH}$ includes noncommutative effects, which are
relevant at short distance comparable to $\sqrt{\theta}$~
\cite{SS}.
 As is shown in Fig. 2, the temperature of the NBH grows
during its evaporation until it reaches to the maximum value
$T_{NBH}=T_m=0.015$ at $r_s=r_m=4.76(M_m=2.4)$ and then falls down
to zero at $r_s=r_0$ which the extremal black hole appears with
$T_{NBH}=0$. As a result, the  noncommutativity restricts
evaporation process to a planck-size remnant, similar to the GUP
inspired black hole~\cite{ACS}. In the region of $r<r_0$, there is
no black hole for $M<M_0$ and thus the  temperature can not be
defined. For $M>M_0$, we have the inner horizon but the observer
at infinity does not recognize the presence of this horizon. Hence
we call this region as the forbidden region.
\begin{figure}[t!]
   \centering
   \includegraphics{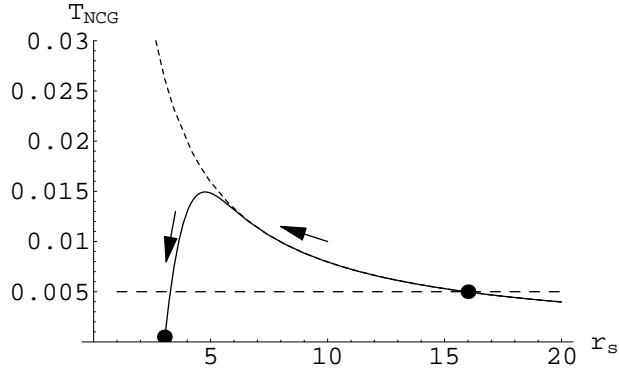}
\caption{The solid line:  temperature $T_{NBH}$ as a function of
the black hole radius $r_s$. The dashed line is the Schwarzschild
case. The horizontal dashed line denotes the initial temperature
$T_i=0.005 (\bullet)$ and $T=0(\bullet)$ represents the end point
for an evaporation process ($\to$). } \label{fig2}
\end{figure}

The entropy of the NBH can be obtained using the
relation\footnote{In deriving Eq.(\ref{entropy}), we use the
first-law of thermodynmics $dM=T_{NBH}dS_{NBH}$. The issue is the
lower bound of the integral. As was shown Fig. 1, we have two
branches for $M>M_0$. The inner branch describes the inner
cosmological horizon, while the outer branch shows the evolution of
outer event horizon. We note that the first-law of thermodynmics
holds for the outer horizon, because the observer at infinity does
not recognize the inner horizon which is beyond the outer horizon.
Furthermore, we have an unphysical negative tempearture $T_{NBH}<0$
for $r_s<r_0$ \cite{GM}. Hence the proper lower bound is not $M=0$
but $M=M_0$.}
\begin{eqnarray}
\label{entropy}
S_{NBH}(r_s)&=&\int_{M_0}^M
\frac{dM'}{T_{NBH}(M')}=\int_{r_0}^{r_s}
\frac{1}{T_{NBH}(r')}\Bigg(\frac{dM'}{dr'}\Bigg)dr'.
\end{eqnarray}
The numerical result of this integration is shown in Fig. 3. In
this case we have zero entropy for the extremal black hole at
$r_s=r_0$. On the other hand, we have the area-law behavior of
$S_{BH}=\pi r_s^2$ for the Schwarzschild black hole.
\begin{figure}[t!]
   \centering
   \includegraphics{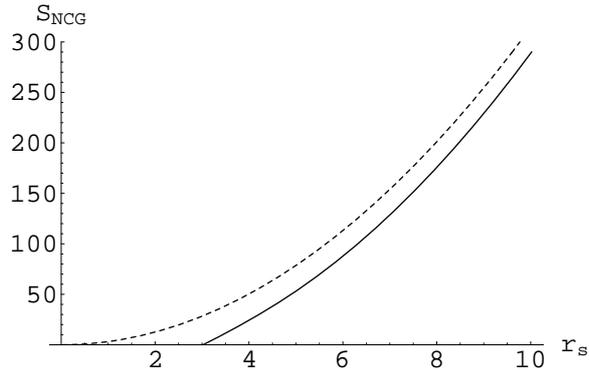}
\caption{The solid line:  entropy $S_{NBH}$ as a function of the
black hole radius $r_s$. The dashed line is the Schwarzschild
case. } \label{fig3}
\end{figure}

In order to check the thermal stability of the NBH, we have to
know the heat capacity.  The heat capacity of the NBH is given by
\begin{equation}
C_{NBH}(r_s)=\frac{dM}{dT_{NBH}}=\Bigg(\frac{dM}{dr_s}\Bigg)\Bigg(\frac{dT_{NBH}}{dr_s}\Bigg)^{-1}
\end{equation}
and its variation  is plotted in Fig. 4. Here  we find a
 stable region of  $C_{NBH}>0$  at the critical regime.
This means that the NBH is thermodynamically stable in the range of
$r_0< r_s <r_m$. The heat capacity becomes singular at $r_s=r_m$
which corresponds to the maximum temperature $T_{NBH}=T_m$. This
picture is consistent with our expectation\footnote{At this stage,
we raise a question: what extend is it physical that the
thermodynamic process would pass through a point $r_s=r_m$ where the
specific heat goes from being infinitely negative to infinitely
positive and then down to a finite positive?  We may understand this
picture from the analogy of the Hawking-Page phase transition in the
AdS black hole \cite{Hawking1, Hawking2}. In the Hawking-Page
transition, we start with the AdS space. A small black hole appears
with negative heat capacity.  The heat capacity changes from
negatively infinity to  positively infinity at the minimum
temperature which corresponds to the maximum temperature of
$T_{NBH}=T_{m}(M=M_m, r_s=r_m)$ in our model. Then the large black
hole with positive heat capacity comes out as a stable object. For
our noncommutative black hole, we may consider the  thermodynamic
process as the inverse Hawking-Page transition. As a result, the
infinite change at $r_s=r_m$ indicates a thermodynamic behavior of
the gravitating system which appears in the process from the large
black hole to the extremal black hole.}. We also observe that a
thermodynamically unstable region ($C_{NBH}< 0$) appears for
$r_s>r_m$.  As a consistent check, we note that in the Hawking
regime, $C_{NBH}$ is consistent with the specific heat of the
Schwarzschild black hole $C_{Sch}=-2\pi r^2_s$.
\begin{figure}[t!]
   \centering
   \includegraphics{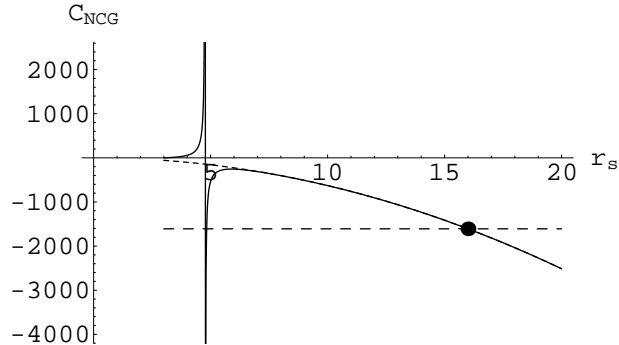}
\caption{The solid line:  heat capacity $C_{NBH}$ as a function of
the black hole radius $r_s$. The dashed curve is the Schwarzschild
case. The horizontal dashed line denotes the initial heat capacity
$C_i=-1608(\bullet)$ for  evaporation.  } \label{fig4}
\end{figure}
\begin{figure}[t!]
   \centering
   \includegraphics{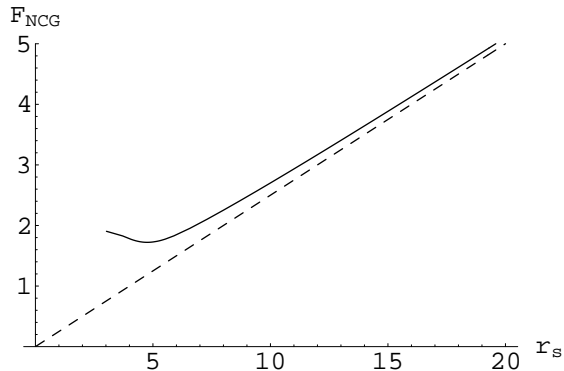}
   \caption{The solid line: plot of the  free energy $F_{NBH}$ as a function of $r_s$.
   The dashed line is the Schwarzschild case $F_{Sch}=M/2=r_s/4$. } \label{fig5}
\end{figure}

Finally, we introduce the free energy as
\begin{eqnarray}
F_{NBH}(r_s)&=&M(r_s)-T_{NBH}(r_s)S_{NBH}(r_s)\nonumber\\
&=& M(r_s)-\frac{1}{4\pi
r_s}\Bigg[1-\frac{r_s^3}{4\theta^{\frac{3}{2}}}\frac{e^{-\frac{r_s^2}{4\theta}}}
{\gamma_s}\Bigg]\int_{r_0}^{r_s}
\frac{1}{T_{NBH}}\frac{dM'}{dr'}dr'.
\end{eqnarray}
Although we do not know its analytic form,  its graph is shown in
Fig. 5 by numerical computations. As expected, the free energy also
has the minimum value at $r_s=r_m=4.76$. We find that two free
energies take the same form in the Hawking regime.  At the critical
regime, two are different. We anticipate that the free energy
$F_{NBH}$ is negative at the critical regime because of positive
heat capacity. However, this is positive.

At this stage we would like to compare our results  with Nozari
and Mehdipour~\cite{NM}. They insisted that there exist negative
temperature, negative entropy, and anomalous heat capacity in the
thermodynamic study of the NBH. They argued that these unusual
features show the failure of standard thermodynamics at the
quantum gravity level. However, these results seems to be
incorrect because they did not consider carefully the fact that
the black hole prescription is meaningful only in the range of
$r_s=r_0(M=M_0)$ to $\infty(M=\infty)$. The observer at infinity
talks about thermodynamics of the outer  horizon. However, he does
not know what happens inside the outer horizon.
 Our  results are correct and consistent
with those for the nonsingular black hole~\cite{Hay}.

Finally we describe  a thermodynamic process, which is closely
related to the evaporation process of the NBH. Let us start with
the black hole with mass $M=M_i> M_0(r_i > r_0)$ in the Hawking
regime. In this case we  sketch the evaporation process ($\to$) by
observing thermodynamic quantities: For a process of $r_i\to r_m
\to r_0$, one has $M_i=8 \to M_m \to M_0$(Fig. 1); $T_i=0.005 \to
T_m \to T_0$ with a sequence of $T_0<T_i<T_m$(Fig. 2).
Interestingly, as is shown in Fig. 4, we have a change of
$C_i(=-1605) \to C_m(-\infty \to \infty) \to C_0(=0.0015).$ Here
we find that the final remnant of extremal black hole  at
$r_s=r_0$ is thermodynamically stable because of positive heat
capacity.

\section{Evaporation of the noncommutative black hole}
We start with the fact that the NBH looks like the regular
solution known as the de Sitter-Schwarzschild black hole. Hence
its causal structure is similar to that of a Reissner-Nordstrom
black hole with the internal singularity replaced by a regular
center. It is known that such a spacetime is unphysical because of
the presence of the Cauchy horizon $r_C$. However, if the NBH
evaporates, the Cauchy horizon is no more real than the event
horizon $r_s$. Actually the evaporating process will terminate at
the point which corresponds to the maximum Cauchy horizon and the
minimum event horizon ($r_C=r_0=r_s$). Hence we do not need to
worry about the presence of the Cauchy horizon. Also it is
interesting to explore the evaporation process of the NBH because
of the absence of a singularity.  We guess that the dynamic
regions are Vaidya-like with the negative-energy flux during
evaporation.

We begin by reexpressing the metric in Eq.(\ref{2eq2})  in terms
of ingoing Eddington-Finkelstein coordinates: $(v,r,\theta,\phi)$.
We introduce the advanced time coordinate \be\label{3eq1}
v=t+r^*,~~r^* \equiv \int^r dr' /F(r'). \ee Here $r^*$ is a
generalization of the tortoise coordinate. Using $dv= dt+dr/F(r)$,
we obtain

\be\label{3eq2} ds^2=-F(r) dv^2 + 2 dv dr +r^2d\Omega_2^2. \ee

Considering the static metric together with  Stefan's law,
the mass dependence of the luminosity  is given by\be \label{3eq3}
L(M)=\sigma A T^4_{NCG} \ee with $A=4\pi r_s^2$ and
$\sigma=\pi^2/60$ for a single massless field with 2 degrees of
freedom~\cite{BR}. Now we are in a position to compute the mass $M(v)$ of
the black hole as seen by a distant observer at time $v$ by
solving the differential equation including first-order in the
luminosity,

\be\label{3eq4} -{d \over dv} M(v) = L(M(v)). \ee

\noindent The improved metric is obtained by replacing the
constant $M$ in $F(r)$ with $M(v)$:

\be\label{3eq5} ds_{NCV}^2=-F(r,v)dv^2 + 2 dv dr
+r^2d\Omega^2,~~F(r,v)=1-\frac{4M(v) \gamma}{r\sqrt{\pi}}. \ee For
$M \gg M_0(\gamma \simeq \sqrt{\pi}/2)$, Eq.(\ref{3eq5}) becomes
the Vaidya metric, which was frequently used to explore the
influence of the Hawking radiation on the Schwarzschild geometry
\cite{bar,his,his2}. It is a solution to Einstein equation
$G_{\mu\nu}=8 \pi T_{\mu\nu}$, where $T_{\mu\nu}$ describes an
inward moving null fluid. In this picture, the decreasing  $M$ is
due to the inflow of negative energy. The metric in
Eq.(\ref{3eq5}) can be regarded as the noncommutativity-corrected
Vaidya (NCV) metric.

It is instructive to ask which energy-momentum tensor $T_{\mu
\nu}$ would give rise to the NCV metric. Computing the Einstein
tensor with Eq.(\ref{3eq5}), one finds that  non-zero components are
\label{3eq6} \ba
&& {T^{v}}_v = {T^r}_r=p_r\label{3eq7},\\
&& {T^r}_v = \frac{{\gamma}
 \dot{M}(v)}{2(\pi)^{3/2} r^2}\label{3eq8},\\
&& {T^{\theta}}_\theta =  {T^\phi}_\phi= p_\bot. \ea
\begin{figure}[t!]
\centering
\includegraphics{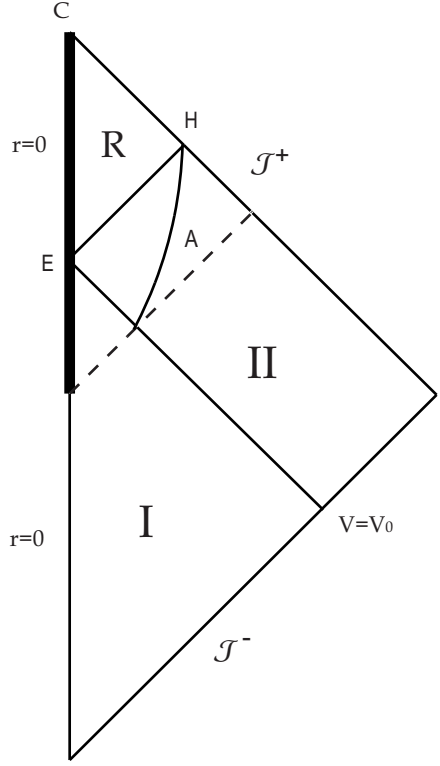}
\caption{The Penrose diagram of the evaporating NBH.   Region $I$ is
a flat spacetime and region $II$ is the evaporating NBH spacetime.
Region $R$ means a planck-size remnant. The line $EH$ is the event
horizon, the line $CH$ is the Cauchy horizon, and the curve $A$ is
the apparent horizon. At $V=V_0$, an imploding null shell exists to
balance the flux of negative energy.}\label{fig6}
\end{figure}
Here the dot denotes the  derivative with respect to $v$. Allowing
for $M(v)\neq$ const, the new feature is  given by a nonzero
component ${T^r}_v$ which describes the inflow of negative energy
into the black hole for $\dot {M}< 0$. This shows pure radiation,
recovering the Vaidya solution for $r^2/4 \gg 1$. In the Vaidya
case, the ingoing radiation creates a singularity. However, the
center remains regular with de Sitter space. This implies that the noncommutative effects protect
the core.

Even though $F(r,v)$ is a complicated function of $r$, both the
early and the late stages of the evaporation process can be
described approximately. We assume that a black hole starts with
$M(v=0)> M_0$ in  the Hawking regime. In this case, we have the
known result of the Schwarzschild black hole

\be \label{3eq9} T_{H} (M)= \frac{1}{8\pi  M},~~L_{Sch}(M) =
\frac{\delta}{M^2}
 \ee
\noindent with $\delta=\frac{\sigma}{256 \pi^3}$. It is easy to
solve the differential equation $-\dot{M}=L(M)$ for this
luminosity. With $M(v=0)=M_i$ the solution takes the form

\be\label{3eq10} M(v)=\Big [M_i^3-3\delta v \Big ]^{1/3}. \ee

\noindent This decreasing  mass  is valid  during the early stage
of the evaporation process,  as long as $M(v)$ is well above the
minimal  mass $M_0$. If one  extrapolates (\ref{3eq10}) to small
mass, one finds $M(v_*)=0$.  This implies that  a final explosion
with $T\rightarrow \infty$ and $L\rightarrow \infty$ occurs, after
a finite time of $v_* = M_i^3 / (3 \delta)$. However, this is not possible
for the NBH. The final stage  of the evaporation process, where
the cold remnant forms, is at the critical regime. It can be described
by those terms which are dominant for $M \to M_0$. Using the
Taylor's expansion of Eq. (\ref{2eq7}) at $M_0$ together with Eq.
(\ref{3eq3}), we obtain the  approximate forms:

\ba && T_{\rm NBH}(M) \simeq \alpha (M-M_0),\label{3eq11} \\
    && L_{NBH}(M) \simeq \beta (M-M_0)^{4}. \label{3eq12} \ea
with $\alpha=dT_{NBH}/dM|_{r_s=r_0}=3242.87$ and $\beta=\sigma A
\alpha^4=2.1 \times 10^{15}$. Solving $-\dot{M} = L(M)$ with Eq.
(\ref{3eq12}), one finds \be \label{3eq13} M(v)-M_{0}=\frac{M_1
-M_{0}}{[1+ 3\beta (M_1-M_{0})^{3}(v-v_1)]^{1/3}} \ee where $v_1$
is a time in the critical region and $M_1=M(v_1)$. For $v \to
\infty$, the difference~\footnote{In the case of quantum-corrected
Newton's constant, it was shown  that the difference of
$(M(v)-M_0)$ vanishes as $v^{-1}$ by using the RG improved Vaidya
metric~\cite{BR}.} of $M(v)-M_0$ vanishes as $v^{-1/3}$. Hence, we
have the late stage  of evaporation: $T_{\rm NBH}(v) \propto
v^{-1/3}$ and $L_{NBH} \propto v^{-4/3}$.  The
noncommutativity-corrected black hole spacetime leads to concrete
predictions on the final state of the evaporation process. We note
again that $M=M_0$ is the mass of a cold remnant, which is  an
extremal black hole with the planck size~\cite{Med}.

Finally, the whole picture of  evaporation process is shown in Fig.
6. Region $I$ is a flat spacetime, while at $V=V_0$ an imploding
null shell is present\footnote{$V$ is the Kruskal advanced time
coordinate, defined as $V =-e^ {-\kappa v}$ with $\kappa=2\pi
T_{NBH}$ the surface gravity of the outer horizon.}. Strictly
speaking, it must have a negative tension in order to balance the
flux of negative energy on its future side \cite{baris}. Region $II$
corresponds to  the evaporating NBH spacetime. The apparent horizon
$A$  is a timelike hypersurface which meets the event horizon at
future null infinity in the Penrose diagram. The null ray of dashed
line, which is tangent to the earliest portion of the apparent
horizon, would have been the event horizon if the black hole were
not radiating. The final remnant $R$ of the NBH is an extremal black
hole whose inner and outer horizons have the same radius of
$r_s=r_C=r_0$.

\section{Discussion and Summary}
There are a lot of  approaches for treating black hole evaporation
process. This process is a quantum gravitational effects and its
understanding provides a suitable framework toward a complete
formulation of quantum gravity. In this work, we have focused on
the thermodynamic approach to the final stage of NBH evaporation.
Our results and corresponding figures indicate stable features
when the mass of the NBH becomes planck scale.  We have obtained a
maximum temperature $T_{NBH}=T_m$ that the NBH can reach before
cooling down to absolute zero ($T_{NBH}=T_0$). Moreover, the
entropy is zero at $M=M_0$ and the heat capacity is positive at
the critical region. These imply that the final remnant is a
thermodynamically stable object. The thermodynamic
 process is closely connected to the evaporation process using the
 NCV metric.  In this case the backreaction effect is
 trivial  because the temperature approaches zero (not divergent) as
$M \to M_0$.

Finally, we have shown that the coordinate coherent approach to
the noncommutativity can cure the singularity problem at the final
stage of black hole evaporation.

\newpage
\section*{Acknowledgement}
This work was supported by the Science Research Center Program of
the Korea Science and Engineering Foundation through the Center for
Quantum Spacetime of Sogang University with grant number
R11-2005-021.  Y.-J. Park was also in part  supported by the Korea
Research Council of Fundamental Science and Technology(KRCF), Grant
No. C-RESEARCH-2006-11-NIMS.

\end{document}